# Evaluation of Energy- and Capacity-Market Revenues from Lithium-ion Battery Systems for Offshore Wind Using Advanced Battery Models

Mehdi Jafari[1], Audun Botterud[1], Apurba Sakti[2*]

[1]Laboratory for Information and Decision Systems, Massachusetts Institute of Technology, Cambridge, MA 02139
[2]MIT Energy Initiative, Massachusetts Institute of Technology, Cambridge, MA, 02139
*Corresponding author: sakti@mit.edu, (+1) 617 715-4512, MIT Building E19-341N, Cambridge, MA 02141

*Abstract*:

Revenue potential from offshore wind and energy storage systems for a Long Island node in the New York ISO (NYISO) is examined using advanced lithium-ion battery representations. These advanced mixed-integer-linear battery models account for the dynamic performance, as well as the degradation behavior of the batteries, which are usually not accounted for in power systems models. Multiple hybrid offshore wind and battery system designs are investigated to examine the impact of locating the battery offshore versus locating it onshore. For the examined systems, we explore different battery usable state-of-charge (SOC) windows, and corresponding dispatch of the battery to maximize energy- and capacity-market revenues. The impacts of variability of offshore wind output along with energy- and capacity-market prices are evaluated using publicly available data from 2010 to 2013. Locating the battery onshore resulted in higher revenues. For 2013, results highlight that without accurate battery representations, models can overestimate battery revenues by up to 155%, resulting primarily from degradation-related costs. Using advanced algorithms, net revenue can be increased by 29%. Results also indicate that wider useable SOC windows could lead to higher net revenues from the energy market, due to higher arbitrage opportunities that compensate for any additional degradation-tied costs in higher DODs. The added value of a MWh of energy storage varies from $2 to $3.5 per MWh of wind energy, which leads to a breakeven cost range of $50-$95 per kWh for the battery systems studied. As such, energy- and capacity-market revenues were found to be insufficient in recovering the investment costs of current battery systems for the applications considered in this analysis.

*Highlights*:

- The battery system, when located onshore, yielded higher revenues
- Without proper battery representations revenues are overestimated by 155%
- Using advanced algorithms, net revenue can be increased by 29%
- Energy- and capacity-market revenues not enough for the battery to breakeven
- Battery breakeven costs found to be in the range of $50-95 per kWh
- Wider SOC-windows for batteries can be more economical in some cases

## 1. Introduction

Decarbonizing the electricity sector by increasing the capacity of renewables in the generation mix is one of the main pathways for reducing greenhouse gas (GHG) emissions [1]. Within renewable energy technologies, offshore wind is expected to have a promising future, due in part to significant lowering of costs in recent years [2]. Fixed bottom wind turbines at capacity-weighted average capital costs of $4,350 per kW in 2018, globally, have experienced a 45% drop in costs since 2015, while floating bottom turbines are the more expensive option at $5,605 per kW but are also expected to become more economical as the technology matures [3]. These capital costs put these turbines within striking distance of other technologies; the levelized cost of energy (LCOE) from fixed bottom offshore wind is reported to be as low as $92-98 per MWh in 2018 [4-5], while that of the floating bottom systems is at $175 per MWh [3], compared to LCOE values of $14-47 per MWh [5] from onshore wind and $32-41 per MWh [5] from utility-scale solar. In the United States, growth has been limited to date with the existence of just one offshore wind farm of 30MW. However, that is expected to change with investments of over $68 billion in the pipeline from about 17GW of planned offshore wind projects [6]. Globally, grid-tied offshore wind capacity additions in 2018 reached almost 4.5GW, which is 15% higher than in 2017, with Chinese deployments tripling to 1.6GW during that year [7].

*Nomenclature:*

**Indices**
| | |
|---|---|
| $t$ | time (hr), $t = 1, \ldots, T$ |
| $d$ | time (day), $d = 1, \ldots, D$ |
| $k$ | SOC levels, $k = 1, \ldots, K$ |
| $l$ | pieces of linearized discharging loss curves, $l = 1, \ldots, L$ |
| $n$ | pieces of linearized charging loss curves, $n = 1, \ldots, N$ |
| $N$ | Indicator for onshore battery variables |
| $F$ | Indicator for offshore battery variables |

**Parameters**

*Market data:*
| | |
|---|---|
| $\pi_e(t)$ | day-ahead market prices ($/MWh) |
| $\pi_c(d)$ | capacity market prices ($/MW-day) |

*Offshore wind:*
| | |
|---|---|
| $P_W(t)$ | offshore wind output power (MW) |
| $C_W$ | offshore wind capacity (MW) |
| $Cr_W$ | offshore wind capacity credit (%) |

*Cable:*
| | |
|---|---|
| $\gamma_{cab.}$ | cable investment cost ($/MW) |
| $A_{cab.}$ | cable's annuity factor |
| $\eta_{cab.}$ | cable efficiency |
| $\eta_{pl}$ | onshore power line efficiency |

*Battery:*
| | |
|---|---|
| $\gamma_b$ | replacement cost ($/MWh) |
| $\gamma_{b,VOM}$ | variable O&M cost ($/MWh) |
| $C_r$ | rated capacity (MWh) |
| $S_{up}$ | upper limit of SOC |
| $S_{dn}$ | lower limit of SOC |
| $\beta(k)$ | SOC bins for loss curves |
| $p_d(l)$ | discharging losses curve pieces |
| $b_d(l)$ | slopes of discharging losses curve pieces |
| $p_c(n)$ | charging losses curve pieces |
| $b_c(n)$ | slopes of charging losses curve pieces |
| $P_d^{Max}$ | maximum discharging power (MW) |
| $P_c^{Max}$ | maximum charging power (MW) |
| $C_{int.}$ | initial energy level (MWh) |
| $Q_{int}$ | initial capacity (MWh) |
| $EOL$ | end of life criteria (remaining capacity) |
| $L_{cal}$ | calendar life (day) |
| $L_{cyc}(DOD)$ | cycle life |

**Decision variables**
| | |
|---|---|
| $E_S(t)$ | sold energy (MWh) |
| $E_P(t)$ | purchased energy (MWh) |
| $P_d(t)$ | discharged power (MW) |
| $P_d^{loss}(t)$ | discharged power losses (MW) |
| $\alpha_d(t, SOC)$ | maximum discharge power (MW) |
| $w_d(t, l, k)$ | discharged power linearized pieces (MW) |
| $P_c(t)$ | charged power (MW) |
| $P_c^{loss}(t)$ | charged power losses (MW) |
| $\alpha_c(t, SOC)$ | maximum charge power (MW) |
| $w_c(t, n, k)$ | charged power linearized pieces (MW) |
| $P_{cW}(t)$ | charged power directly from the wind (MW) |
| $P_s(t)$ | wind power directly sold to the grid (MW) |
| $P_{cab.}(t)$ | cable power (MW) |
| $C_{cab.}$ | cable capacity (MW) |
| $P_{curt.}(t)$ | curtailed wind power (MW) |
| $C(t)$ | battery energy level (MWh) |
| $S(t)$ | battery SOC (p.u.) |
| $C_{act}(d)$ | battery's actual capacity (p.u.) |
| $Q_{cal}(d)$ | capacity fade due to calendar degradation |
| $Q_{cyc}(d)$ | capacity fade due to cycling degradation |
| $B(t)$ | binary variable to control charge/discharge |
| $U(t, k)$ | binary variable for SOC curves selection |

**Other**
| | |
|---|---|
| $R_{C,W}$ | capacity market revenue for offshore wind |
| $R_{C,B-ISO}$ | capacity market revenue for ISO-managed BESS |
| $R_{C,B-self}$ | capacity market revenue for self-managed BESS |
| $k(d)$ | number of hours in each day that SOC is higher that its lower limit |

Increasing offshore wind deployments will further elevate the concerns that are emerging regarding the challenges of grid integration and system reliability with the rise of variable renewables. The materiality of these concerns varies from system to system, however, one mechanism to mitigate any issues related to variability and uncertainty is energy storage [8]. Advanced energy storage technologies such as batteries, can provide the grid with the added flexibility needed to reliably accommodate much higher levels of variable renewable generation. This potential for a strong synergistic relationship between storage and renewables is expected to result in the deployment of significant amounts of advanced storage assets across many power systems over the coming years, and indeed this dynamic has already started. In the US alone, the first quarter of 2017 witnessed deployment of 71MW of battery energy storage projects, a 276% increase over the first quarter of 2016. In 2018, energy storage deployments, at 777MWh, was an 80% year-on-year increase compared to 2017 [9]. Such levels of deployment can be expected to be repeated again in 2019 as the economics of energy storage systems continue to improve. Prices of lithium-ion batteries, the dominant battery technology, has decreased from an average $900/kWh in 2009 [10] to $209/kWh in 2018, at the pack level [11].

Now, multiple studies have investigated the economic potential of offshore wind both with and without an accompanying energy storage system [4, 12-14]. Mills et el. [12] developed a model to study the profitability of offshore wind in the US using historical data and concluded that the revenue potential varies significantly with location. Beiter et al. [4] calculate and compare both the levelized cost of energy (LCOE) for projects in the northeastern US, as well as, the levelized revenue of energy (LROE) for offshore wind using power purchase agreement (PPA) data between Massachusetts distribution companies and Vineyard Wind LLC. They found that LCOE estimates of $120-160/MWh for offshore wind projects in the northeastern US exceed the calculated LROE of $98/MWh using Vineyard Wind's power purchase agreements

(PPAs) in the US. Beiter et al. [4] hypothesize that this discrepancy between LCOE and the LROE, which in a perfectly competitive market can be expected to be equal, could result from a range of factors including the US's nascent market benefiting from cost-reduction trajectories in Europe. Other studies have investigated the profitability of energy storage systems at the grid-level under different market conditions [15-19]. He et al. [15] develop optimal bidding strategies for battery energy storage systems (BESSs) to participate in the energy market while accounting for the life of the battery. Bradbury et al. [16] evaluate the economic viability of BESSs for arbitrage in real-time markets by using a simple linear BESS model. In a more comprehensive study, Davies et al. [17] compare the revenue potential of different battery chemistries including lithium-ion, nickel-cadmium, and sodium-ion using a linear battery model with a constant battery roundtrip efficiency. Wankmüller et al. [18] develop advanced battery models by accounting for the impact of battery degradation. In another study, Sakti et al. [19] propose enhanced lithium-ion battery models that consider variable efficiencies and maximum power limits as a function of the battery's state-of-charge (SOC), however, Sakti et al.'s models did not account for the impact of the battery's degradation.

Most existing studies focus either on offshore wind or on BESSs separately, with the ones that consider both, relying on general linear models of BESSs, which ignore its non-linear performance and degradation behavior. A gap exists in the literature when it comes to the evaluation of the economic viability of offshore wind connected BESSs that consider better representations of battery behavior to estimate the added value from the BESS more accurately. In this paper, we develop models to fill this void. We build up on Sakti et al.'s [19] prior work on enhanced battery representations by modifying the BESS's dynamic efficiency representation and accounting for both calendar- and cycle-life degradations. These advanced battery models are then used to evaluate BESSs for offshore wind applications for different system designs and markets using 2010-2013 wind and market data specific to a NYISO node. System designs include the location of the battery to investigate whether locating a battery offshore offers any benefits to siting it onshore, particularly from lower submarine cable costs by operating a lower-capacity cable at higher capacity factors. Results provide useful insights into optimal system design and the economic viability of different BESS solutions. The rest of paper is structured as follows: section 2 presents the different designs for offshore wind connected BESSs that we investigate, section 3 discusses the methodology and problem formulation in MILP form, section 4 highlights the assumptions of this study, section 5 presents the results and discussion, and the summary and conclusions are presented in section 6.

## 2. OFFSHORE WIND & BESS DESIGN CONFIGURATIONS

We investigate four different design configurations for the offshore wind farm: i) the offshore wind farm with no BESS,
ii) BESS located onshore, iii) BESS located offshore, and iv) a hybrid system with BESSs both onshore and offshore. The general schematic of these designs is shown in Fig. 1. The different locations for the BESSs is investigated to evaluate the impact on the overall profitability of the combined system.

In the studied system, the BESSs are primarily charged using offshore wind energy. The wind energy can also be directly delivered to the grid through the cable and onshore power line. In the event that the generated wind power is more than the cable's rated capacity and the BESS's charging power limit, the additional power is curtailed. The BESS can also be charged from the grid when the wind generation is not high enough and the power prices are low. This energy from the BESS can then be discharged to the grid at a later time. Hence, in the designed system, the BESS can be used to time-shift the generated wind energy or participate in direct energy arbitrage from the grid. The BESS can also participate in the capacity market, if the regulations of the host ISO allow for such participation. While the onshore and offshore BESSs can of course be from different technologies, in this work, we consider a Li-ion battery, more specifically a Lithium Nickel Manganese Cobalt Oxide (NMC) chemistry.

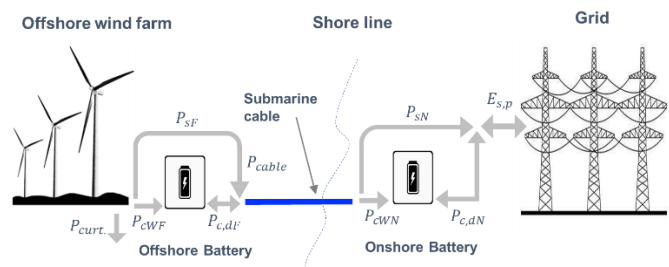

Fig. 1. Offshore wind connected battery system configurations: sold and purchased energy ($E_{s,p}$), energy exchange of onshore (*N*) and offshore battery (*F*) with the grid ($P_{c,dN,F}$), charged energy from wind generation ($P_{cWN,F}$), wind energy directly sold to the grid ($P_{sN,F}$).

## 3. METHODOLOGY

Economic evaluation of the mentioned systems and its revenue estimation in different markets are performed through an optimization model using a mixed integer linear programming (MILP) formulation. Decision variables include the dispatch of the BESS, wind curtailment, and cable sizing. The objective function maximizes the revenue of participation in the energy market, while considering the operational costs and constraints. Traditional modeling methods of BESS in power system problems usually assume fixed roundtrip efficiencies and/or fixed rated power for the batteries [16-17]. However, in reality, the BESS's charging and discharging efficiencies vary by the output power and SOC, while its maximum power, also, is a function of SOC [19-20]. More importantly, neglecting the degradation of the battery has been shown to impact its optimal cycling profile and significantly overestimate the revenue potential of the BESS [18]. As such, the analysis presented in this study incorporates an enhanced

BESS model, which accounts for the battery's dynamic efficiency and maximum cycling power, in addition to its calendar- and cycle-degradation. However, due to non-linear nature of the efficiency (losses) and maximum power in different SOCs, we implement a piecewise linearization method to be able to capture these dynamic behaviors in MILP form. Capacity fade also behaves non-linearly with respect to time and cycling; however, we incorporate a linear model for the same.

The optimization formulation of the problem considers the offshore wind connected BESS to be participating in the energy market and considers its operational costs. The objective function maximizes energy revenues. Mathematically,

$$\max_{E_S E_P} R = \sum_{d \in D} \sum_{t \in T} [\pi_e(t)(E_S(t) - E_P(t)) - \gamma_{b,VOM} \times P_d(t)] \\ -(1 - C_{act}(D)) \times \gamma_b \\ -C_{cab.} \times \gamma_{cab.} \times A_{cab.} \quad (1)$$

where, $\pi_m$ is the day-ahead energy price in a specific market at time $t$, and $E_P$ and $E_S$ are the purchased and sold energy from and to the gird, $t$ and $d$ define the time indices of hours and days. $\gamma_{b,VOM}$ is variable O&M cost of the battery per MWh, $\gamma_b$ is the battery replacement cost per MWh, $C_{act}$ is the actual capacity of the battery after degradation, and $C_{cab.}$, $\gamma_{cab.}$ and $A_{cab.}$ are the capacity, investment cost per MW, and annuity factor, respectively, of the submarine cable. Note that in addition to the BESS degradation cost, three other cost elements have been considered for the battery: investment, fixed O&M, variable O&M costs. As the BESS in this problem is a price taker, optimizing its size does not provide much insight; either the battery investment is profitable based on its investment cost and the optimum solution is the maximum allowable battery capacity, or it is not profitable and the solution suggests zero battery installation. Therefore, the battery size and its investment cost are considered as exogenous variables. The fixed O&M cost is tied to the battery size and it is therefore constant for a given battery size. However, the battery's variable O&M and degradation costs are optimization variables. The constraints of this problem change for different BESS locations of onshore, offshore and hybrid design, as outlined below.

### 3-1. Onshore BESS

In the onshore BESS configuration, it is assumed that the generation from the wind farm is either transferred through the cable to the shore or curtailed, if it exceeds the optimum capacity of the cable. The energy transferred to the shore can either be sold directly to the grid, or stored in the onshore BESS. In this configuration, energy balance constraints will be as (2)-(5).

$$E_S(t) = \eta_{pl} \times (P_d(t) - P_d^{loss}(t) + P_s(t)) \quad \forall t \quad (2)$$

$$E_p(t) = (P_c(t) + P_c^{loss}(t) - P_{cW}(t))/\eta_{pl} \quad \forall t \quad (3)$$

$$P_W(t) = P_{cab.}(t) + P_{curt.}(t) \quad \forall t \quad (4)$$

$$P_{cW}(t) = \eta_{cab.} P_{cab.}(t) - P_s(t) \quad \forall t \quad (5)$$

Constraint (2) calculates the sold energy $E_S$ as the discharged power minus its losses ($P_d$ and $P_d^{loss}$), and directly sold wind energy to the grid $P_s$, accounting for the onshore power line efficiency $\eta_{pl}$. (3) accounts for the purchased energy from the grid to charge the BESS in addition to the charging from the wind power, $P_{cW}$. Eqs. (4) and (5) are the energy balance at the offshore and onshore side of the cable, respectively. So, the problem formulation for onshore battery location is objective function (1), subject to (2)-(5) and the battery cycling constraints which will be presented later in section 3-4.

### 3-2. Offshore BESS

For the offshore BESS configuration, it is assumed that the BESS is installed by the wind farm side and combined offshore wind power and battery's discharged power is transferred to the grid through the cable and power line (Fig. 1). Considering this configuration in the problem formulation, the objective function (1) is subject to the energy balance constraints (6)-(9) instead of (2)-(5).

$$E_S(t) = (P_d(t) - P_d^{loss}(t) + P_s(t)) \times (\eta_{cab.} \eta_{pl}) \quad \forall t \quad (6)$$

$$E_p(t) = (P_c(t) + P_c^{loss}(t) - P_{cW}(t))/(\eta_{cab.} \eta_{pl}) \quad \forall t \quad (7)$$

$$P_W(t) = P_{cW}(t) + P_s(t) + P_{curt.}(t) \quad \forall t \quad (8)$$

$$P_{cab.}(t) = (E_S(t) - E_p(t))/(\eta_{cab.} \eta_{pl}) \quad \forall t \quad (9)$$

where, (6) is the sold energy to the grid by discharged energy from the BESS minus its losses and the directly sold wind energy considering the cable and power line efficiency. Constraint (7) calculates the purchased energy to charge the BESS, (8) assures the energy balance on the wind farm side and (9) calculates the cable power.

### 3-3. Hybrid Design

The hybrid design of the system includes both onshore and offshore BESS that can be of different capacities and technologies. To define this configuration, the objective function of (1) will be subject to the energy balance and performance constraints for both the onshore and offshore batteries. Therefore, the BESS cycling constraints will be repeated for both onshore and offshore batteries, and (2)-(5) will be replaced by new energy balance constraints (10)-(14).

The whole optimization problem will be maximizing (1) subject to two sets of BESS cycling constraints ((15)-(19) and (A.1)-(A.20)) and (10)-(14).

$$E_S(t) = (P_{dN}(t) - P_{dN}^{loss}(t) + P_{sN}(t)) \times \eta_{pl} \quad \forall t \quad (10)$$

$$E_p(t) = ([P_{cN}(t) + P_{cN}^{loss}(t) - P_{cWN}(t)] + [P_{cF}(t) + P_{cF}^{loss}(t) - P_{cWF}(t)]/\eta_{cab.})/\eta_{pl} \quad \forall t \quad (11)$$

$$P_W(t) = P_{cWF}(t) + P_{sF}(t) + P_{curt.}(t) \quad \forall t \quad (12)$$

$$P_{cab.}(t) = P_{dF}(t) - P_{dF}^{loss}(t) + P_{sF}(t) \quad \forall t \quad (13)$$

$$P_{cWN}(t) = \eta_{cab.} P_{cab.}(t) - P_{sN}(t) \quad \forall t \quad (14)$$

In this case, the sold energy is related to the onshore battery discharge and the power coming from the cable as in (10). The purchased energy should consider charging of both onshore and offshore batteries as written in (11). (12) accounts for the energy balance in the wind farm side and (13) and (14) present energy balance in the offshore and onshore ends of the cable.

3-4. BESS Model

Regardless of the BESS location, we include constraints defining its cycling and degradation characteristics in the formulation. In this study, building on our previous work in [19-20], we developed an advanced BESS model accounting for both dynamic efficiency and power limits as well as cycle and calendar degradations. Note that the main constraints reflecting the advanced BESS modeling are discussed in this section, while the more basic constraints are presented in Appendix A.

*3-4-1. Dynamic efficiency and cycling power limits*

To define the BESS cycling in all configurations and to include the charge/discharge powers' limits and SOC window, (15)-(17) are written as follows:

$$C_r \times S_{dn} \leq C(t) \leq C_r \times S_{up} \quad \forall t \quad (15)$$

$$P_d(t) \leq \alpha_d(t, SOC) \quad \forall t \quad (16)$$

$$P_c(t) \leq \alpha_c(t, SOC) \quad \forall t \quad (17)$$

$$P_d^{loss}(t) = \sum_{k \in K} \sum_{l \in L} (b_d(l) \times w_d(t, l, k)) \quad \forall t \quad (18)$$

$$P_c^{loss}(t) = \sum_{k \in K} \sum_{n \in N} (b_c(n) \times w_c(t, n, k)) \quad \forall t \quad (19)$$

where, constraint (15) limits the BESS charge and discharge level to predefined minimum and maximum SOCs, $S_{dn}$ and $S_{up}$ respectively. (16) and (17) assure that the charge and discharge powers are limited to their maximum value at each SOC. $\alpha_d$ and $\alpha_c$ are defined to regulate the maximum charging and discharging powers in different SOC levels. For instance, Fig. 2 shows how maximum cycling power changes for different SOCs. Also, we calculate charging and discharging losses $P_c^{loss}$ and $P_d^{loss}$ from dynamic efficiency profiles for each SOC level for selected batteries. Fig. 2 shows the resulting power losses for 30%, 50%, and 70% SOCs from dynamic efficiency data. Comparing these three curves to the constant efficiency line (black line) reveals that the losses calculation with constant efficiency can lead to underestimation or overestimation error for losses at different output powers. To avoid this error and have more accurate estimation of the cycling losses, we have used dynamic efficiency curves by piecewise linearization and integer variables. Constraints (18) and (19) calculate the dynamic discharge and charge losses using calculated power curve pieces $w_{d,c}$ and their slopes $b_{d,c}$. The detailed constraints to formulate the selection of SOC curves and power losses and power limits calculations are presented in Appendix A. Solving (1) subject to energy balance constraints in each configuration and the BESS cycling constraints will optimize the operation of the battery based on the wind generation and the energy price signals.

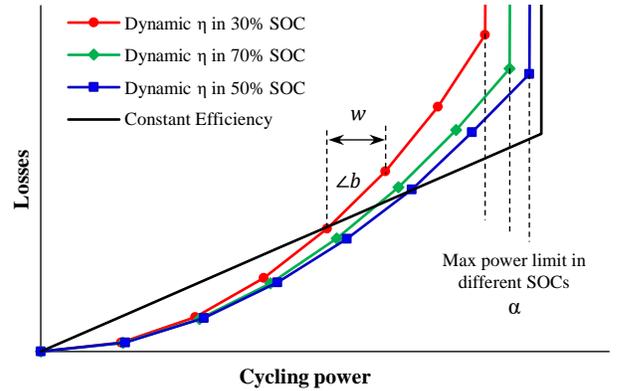

Fig. 2. Constant vs. dynamic maximum cycling power and efficiencies in different SOCs and powers

*3-4-2. Cycle and calendar degradations*

Degradation considered in this study pertains only to the capacity fade occurring in the battery cells, which results from two sources: i) calendar aging, which is a function of time, regardless of the battery use, and ii) cycle aging, which happens by charging and discharging of the battery. For both calendar and cycle degradation mechanisms, temperature plays a key role, while in the case of cycle aging, the depth of discharge (DOD), C-rate and the average SOC are additional contributors [21-22].

From the power system application point of view, the degradation effect can be incorporated into the optimization problem in two ways: reducing the cyclable capacity of the BESS in the problem constraints; and, defining a degradation penalty in the objective function to avoid excessive cycling. Both these aspects are implemented in the model developed in this study to analyze the effect of degradation in the operational and financial results of the BESS. In the formulation, we assume that the battery capacity due to the calendar aging declines linearly by time as shown in (20).

$$Q_{cal}(d) = Q_{int}\left(1 - (1 - EOL) \times \frac{d-1}{L_{cal}}\right) \quad (20)$$

where, $Q_{cal}$ and $Q_{int}$ are the battery's capacity after calendar degradation and its initial capacity, respectively. *EOL* is the end of life criteria, which is the remaining capacity after degradation, $d$ is the number of days and $L_{cal}$ is the calendar life of the battery in days. The degradation rate in this equation is a function of number of days, *EOL* criteria and calendar life of the battery. To account for the battery's cycle degradation ($Q_{cyc}$), the method is to count the number of equivalent full cycles and compare it to the cycle life ($L_{cyc}$) of the battery. In each day, the charging power $P_c$ is summed up and divided by the battery's rated capacity ($C_r$) as follows:

$$Q_{cyc}(d) = Q_{int} \qquad d = 1 \quad (21)$$

$$Q_{cyc}(d) = Q_{int}\left(1 - (1 - EOL) \times \frac{\sum_{t \in 24(d-1)} P_c(t)}{C_r L_{cyc}(DOD)}\right) \quad d \geq 2 \quad (22)$$

To include the effect of DOD in the battery cycle life, we use empirical data of cycle life in different DODs as shown in Fig. 3 [23]. For each DOD based on the operating SOC window, the cycle life is calculated before the optimization and used to estimate the cycle degradation of the battery in the optimization. Therefore, we change $L_{cyc}(DOD)$ in constraint (22) for different DODs.

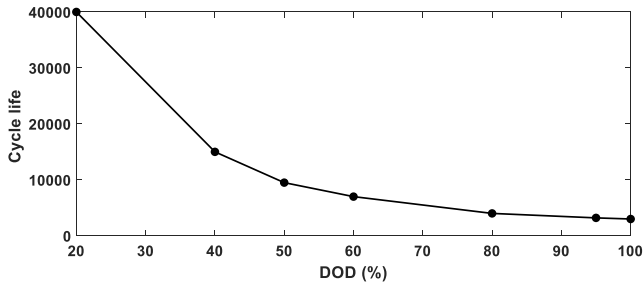

Fig. 3. Battery's cycle life vs. DOD [23]

To implement the degradation in the optimization problem constraints, actual capacity of the battery after cycle degradation is calculated at the end of each day and is used for the cycling limit in the next day. Therefore, constraints (21)-(23) are added to the problem to calculate the actual capacity.

$$C_{act}(d) \leq Q_{cyc}(d) \qquad \forall d \quad (23)$$

Then, the battery's rated capacity in (15) is replaced by its actual capacity from (23) as shown in (24).

$$C_{act}(d) \times S_{up} \leq C(t) \leq C_{act}(d) \times S_{up} \qquad \forall t, d \quad (24)$$

Also, the power rating of the battery is considered to decline linearly with the capacity fade. Therefore, maximum charging and discharging powers are fading by the degradation and a factor of $C_{act}(d)/C_r$ is multiplied to the maximum power constraints of (A.9)-(A.12). As an example, (A.9) will be replaced by (25).

$$\alpha_d(t) \leq \left[\frac{C_{act}(d)}{C_r}\right] \times \sum_{l \in L} p_d(l) + M \times (1 - U(t,k)) \quad \forall t, k, d \quad (25)$$

Note that introducing degradation penalty in the objective function will prevent excessive cycling of the battery to avoid the penalty. However, the degradation will happen even without cycling due to the calendar aging. Therefore, to make sure that the battery is being cycled enough to cover the calendar aging, we add another constraint to set the lower bound for the degradation as (26). Constraints (20)-(26) are added to the problem formulations in all configurations.

$$C_{act}(d) \leq Q_{cal}(d) \qquad \forall d \quad (26)$$

3-5. Capacity market representation

For the offshore wind, we consider constant daily capacity credit and assume that it receives the capacity credit for all days of year, and for simplicity we do not consider any penalty mechanisms. Therefore, capacity market revenue for offshore wind, $R_{C,W}$ is calculated by (27).

$$R_{C,W} = Cr_W \times C_W \times \sum_{d \in D} \pi_c(d) \quad (27)$$

where, $Cr_W$ is capacity credit as percentage of the installed capacity, $C_W$. $\pi_c(d)$ is the daily capacity price. For the BESS, there are several requirements to participate in the capacity market in NYISO such as minimum injection capacity (0.1 MW), capacity resource interconnection service (CRIS), minimum duration requirement (4 hours) and availability factor [25]. NYISO allows energy storage resources to derate their capacity to meet the 4-hour duration requirement. Also, ISO-managed and self-managed BESS receive different capacity credits. For the ISO-managed BESS, we calculate the capacity revenue as follows:

$$R_{C,B-ISO} = \frac{1}{4}(S_{up} - S_{dn}) \times C_r \times \sum_{d \in D} \pi_c(d) \quad (28)$$

which derates the capacity credit by the duration and SOC window and gets the capacity credit for all days. The ISO-managed BESS receives capacity credit regardless of its SOC, while self-managed BESSs do not receive capacity credit for the times that SOC has reached its lower limit (i.e. the BESS cannot be discharged). Therefore, capacity revenue for the self-managed BESS has another factor, adjusting its availability based on SOC level as (29).

$$R_{C,B-self} = \frac{1}{4}(S_{up} - S_{dn}) \times C_r \times \sum_{d \in D} \frac{k(d)}{24} \times \pi_c(d) \qquad (29)$$

where, $k(d)$ is number of hours in each day that SOC is higher that its lower limit.

## 4- ASSUMPTIONS

We consider a hypothetical case with 10 MW installed capacity of offshore wind and evaluate the added value of 1 MWh BESS as the baseline in this study. Fig. 4 shows hourly wind capacity factor data for January 2013 in a location outside Long Island within the NYISO area. We assume that the generation from the offshore wind farm can be directly sold to the grid or charge the batteries. Moreover, there is no penalty for charging from the grid, or renewable energy credit for charging from wind energy. However, the efficiency of the onshore power line to the nearest power station naturally prioritizes charging from the wind energy. Table 1 shows the fixed parameters used for the BESS. The battery chemistry used in this study is Lithium Nickel Manganese Cobalt Oxide (NMC) UR18650E. The efficiency data in different power and SOCs are calculated from the manufacturer data [25]. The battery related costs for onshore and offshore configurations are different. So, for the sake of comparison in this study, we consider offshore battery costs (investment and O&M) to be 20% higher than onshore battery. The BESS investment cost varies with energy to power ratio (durations). For 1-hour battery in this study, the investment costs are $165/kWh and $530/kW [26]. When it comes to the cost of the submarine export cables, values are largely dependent on AC versus DC options, the transmission distance, capacity, and water depths [27]. NREL reports that the cost of two three-phase cables for a 250MW wind plant to be as high as $80-100 million including burial costs at depths of up to 100 meters and a transmission distance of 30 kilometers [28]. In our analysis, we assume a conservative cost-estimate of $125,000 per MW of HVAC cable capacity for our hypothetical 10MW wind farm located at a distance of 20km from the shore. The distance of 20km was chosen based on distribution-data of actual wind farm installations in Europe [28].

For energy market revenue evaluation, we select the Long Island area in New York ISO (NYISO) and use day-ahead market prices from 2010 to 2013 for a coastal node (#23522) as input for the optimization. Sample price data for 2013 are shown in Fig. 5. We assume that the energy transactions with the grid does not change the local energy price and that offshore wind connected BESS is a price-taker. Capacity market prices per kW-year from NYISO for the same time period (2010-2013) are translated to $/MW-day (Fig. 5) and used for the capacity market revenue evaluation. For offshore wind, the capacity credit is 38% of its installed capacity for both summer and winter seasons [29].

Table 1. BESS parameters used in the analysis

| Parameter | Value |
|---|---|
| Capacity | 1 (MWh) |
| Initial charge | 50 (%) |
| Maximum charge | 99 (%) |
| Minimum charge | 1 (%) |
| Maximum power | 1.337 (MW) |
| Maximum/Minimum SOC for baseline | 85/30 (%) |
| Capital recovery period | 10 (years) |
| Interest rate | 0.07 |
| Investment cost ($/kWh) | 165 |
| Fixed O&M cost ($/kW-year) | 8 |
| Variable O&M cost ($/MWh) | 2.3 |

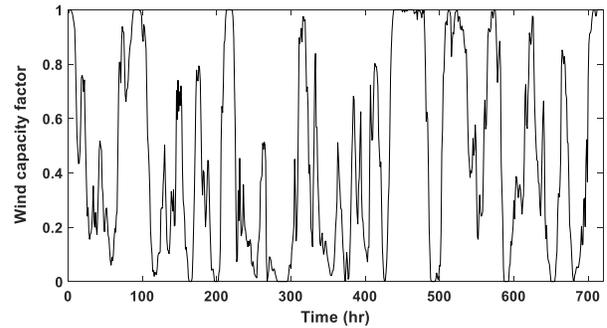

Fig. 4. Sample hourly wind data for one month (January 2013)

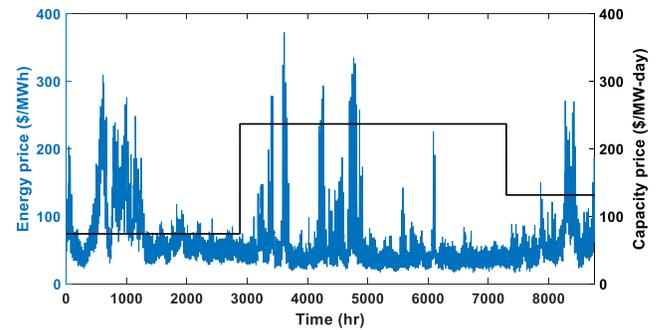

Fig. 5. Energy price for node (#23522 on Long Island, NY) and capacity price for the Long Island zone (NYISO) 2013

## 4- RESULTS AND DISCUSSION

In this section, we first show how the accuracy of the BESS model is critical to better decision-making tools and why the advanced models presented in this study offer advantages over simpler modeling approaches. Subsequently, we present findings from different case studies analyzing impacts of battery location, DOD, and participation in energy and capacity markets.

### 4-1. Battery Modeling Accuracy

As mentioned in section 3, most BESS models in power system analyses assume constant efficiency and constant power limits for all SOCs and discharge rates, without accounting for cycle or calendar life. As expected, these simplifications lead to inaccuracies in modeling the battery cycling profiles, as well as the calculated BESS revenues from the electricity market. Fig. 6 highlights the varying charge and discharge cycles of the BESS using four different models for 2013 data. In the basic model, the optimization is based on the constant efficiency of the BESS and therefore the battery is cycled with the static maximum power as long as the price gap is enough to cover the roundtrip losses. When the model accounts for the dynamic efficiency of BESS, optimization results show lower power charging and discharging profiles to operate the battery in higher efficiencies. The last 36 hours in Fig. 6 are zoomed to better illustrate this difference between basic and dynamic efficiency models. Also, due to the more accurate accounting of losses, the dynamic efficiency model can identify smaller arbitrage opportunities which leads to higher number of equivalent full cycles (Table 2). Note that, as the basic and dynamic efficiency models do not consider the degradation cost, they both lead to excessive cycling of the battery.

Introducing the degradation cost as a cycling penalty in the objective function reduces the number of cycling events. In this case, cycling is limited to a higher price differences that can compensate both the losses and degradation costs due to the cycling. For instance, in the sample period in Fig. 6, the battery is not discharged in 36 hours due to lower prices and is only being charged during the lowest prices of the sample period (see zoomed inset for "cycling degradation" in Fig 6). Overall, optimal arbitrage cycling based on the assumptions in this study tends to have much fewer battery cycles when the cycling degradation is included in the optimization, as also shown in Table 2. However, when the calendar degradation is included as a lower boundary constraint for the capacity fade, the number of cycling events goes up to use the maximum arbitrage opportunity to compensate for the inevitable calendar degradation.

Table 2. Annual equivalent full cycles with different modeling approaches (2013 data)

|  | Basic model | Dynamic efficiency | Cycling degradation | Calendar degradation |
| --- | --- | --- | --- | --- |
| **Cycles per year** | 226 | 241 | 104 | 130 |

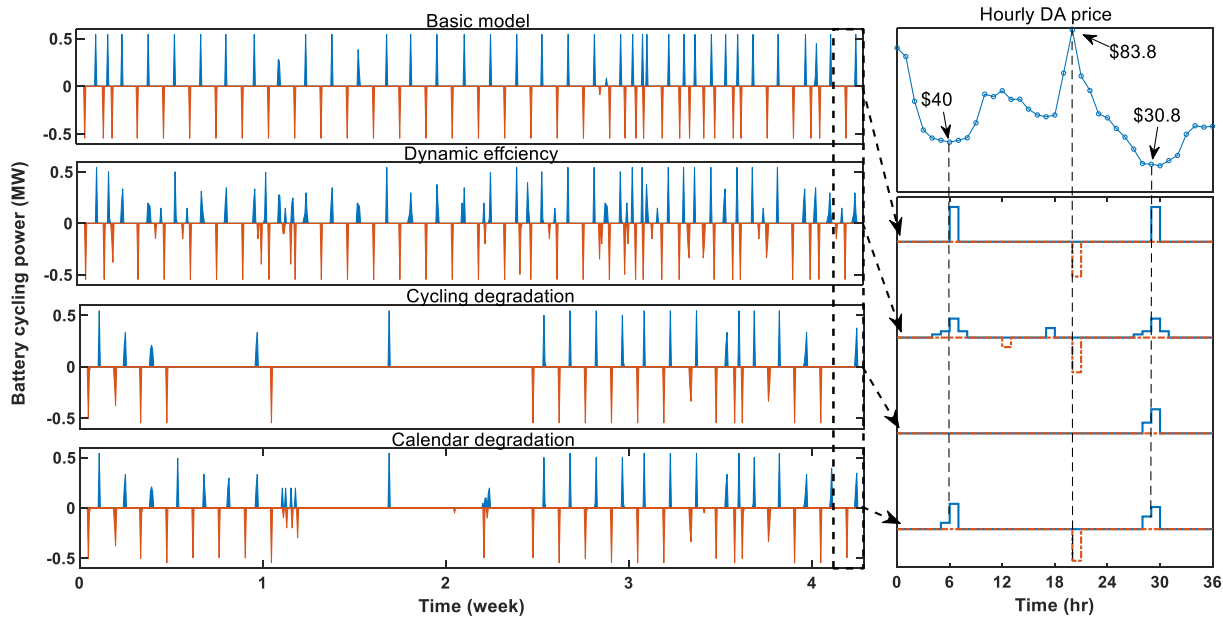

Fig. 6. Simulated battery cycling using different modeling approaches (2013 data)

These different cycling profiles resulting from varying degrees of model sophistication yield different estimates of net battery revenue, which is energy market revenue minus the degradation and variable O&M costs as shown in Fig. 7. In the *basic* model, the total revenue from the energy market is maximized without accounting for the degradation costs, thereby overestimating the battery's economic viability. The *dynamic efficiency* model has 2.4% higher total revenue compared to the basic model due to more accurate estimation of losses and capturing smaller arbitrage revenues, although post-optimization calculation of the degradation costs due to over cycling of the battery shows that it does not necessarily lead to higher net revenues. The *cycling degradation* model has the highest net revenue estimation; however, it does not account for the time-dependent calendar degradation that will happen even if the battery is not cycled. The *calendar degradation* model, by building up on the *dynamic efficiency* and the *cycling degradation* models, is the most comprehensive of all and represents the physical performance of lithium-ion batteries of NMC chemistry more accurately. As such, revenues estimated using this model can be expected to be more reliable. Results using this advanced model show that almost 50% of revenues generated from the energy market by a BESS can be expected to go towards compensating for the degradation and variable O&M costs of the BESS. Overall, ignoring accurate representations of lithium-ion batteries, particularly their degradation, resulted in a total revenue estimate of $10,675, which is 155% greater than the net revenue calculated using the most advanced *calendar degradation* model of $4,185 for the 1MWh battery for 2013. However, it is entirely plausible that even when using the basic model, the user may account for any degradation-related costs, ex-post, once the revenues have been estimated, and on doing so the net revenue is seen to go down to $3,244 (Figure 7). Using our advanced *calendar degradation* model, this net revenue can be increased by 29% to $4,185 since our algorithm prevents excessive cycling of the battery especially when the degradation costs are greater than the revenues generated.

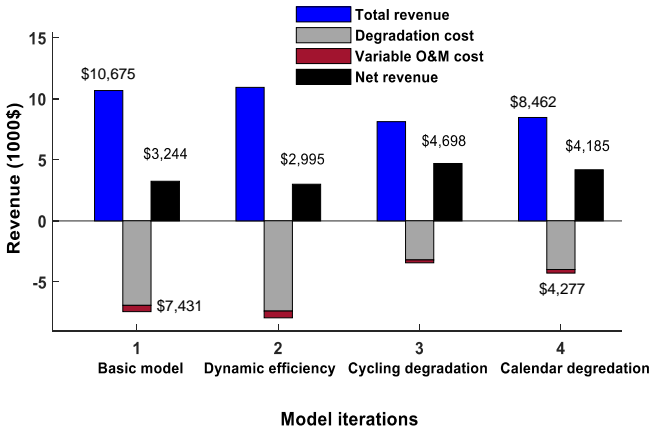

Fig. 7. Revenue and cost comparisons for different modeling iterations for a year for the 1MWh battery using data from 2013

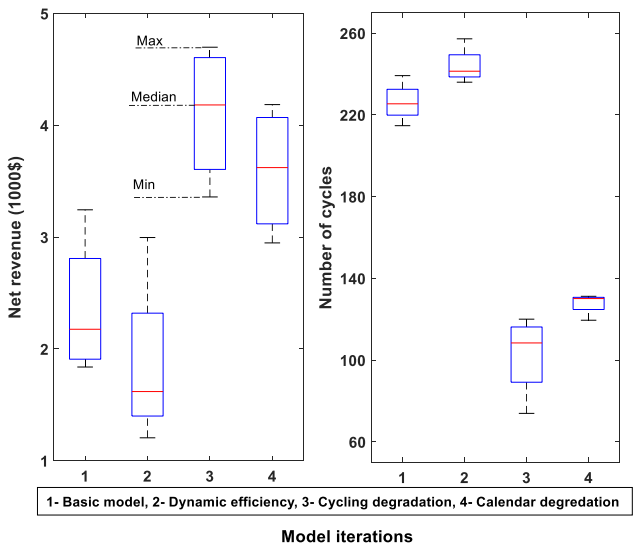

Fig. 8. Box-plot for the net revenue and annual number of cycles in different years using data from 2010-2013 for the 1MWh battery

To examine the robustness of the model's results for different input data, we run four iterations of the model for 2010-2013 wind and price data and present the results in Fig. 8. The figure illustrates that the models' results are robust across wind and price data of different years and follow the same trends as in Fig. 7 and Table 2, indicating the importance of modeling approach in evaluating the BESS revenue in the energy arbitrage market. The results also reveal that the BESS revenue can vary substantially between different years. Using the more sophisticated and accurate *calendar degradation* model, we report our evaluation of energy- and capacity-market revenues from lithium-ion batteries for offshore wind in the next section.

4-2. Revenue in Different BESS Locations

To explore the impact of battery location, we apply the optimization model to onshore BESS, offshore BESS and hybrid designs for the four different years. In the hybrid design, we assume that BESS is split with 50% onshore and 50% offshore. Fig. 9 shows the stacked average (over 2010-2013 data) revenue and cost elements per MWh wind energy in different battery locations. These results reveal that without BESS, the average value of offshore wind is $56 per MWh of wind energy. Adding the 1MWh battery increases this value to $59/MWh. However, the battery associated costs (investment and O&M) and cable investment cost decreases the revenue by 57-68% for different BESS locations. The net revenue of the offshore wind asset per MWh energy (without considering the offshore wind investment and O&M costs) drops from $53 in the case of no BESS to $19, $22 and $25 for offshore, hybrid and onshore BESS cases. Therefore, battery participation in arbitrage and capacity markets only does not compensate its costs, and additional revenue streams, e.g. through other markets such as ancillary services, are needed for the battery to be economically viable under the cost assumptions used in the study.

Table 3 summarizes the breakdown of revenues and costs, and also presents the total value of the battery and its breakeven cost for the three different locations. On average for different BESS locations, 95.6% of the revenue comes from wind's energy and capacity markets and 4.4% is the portion of battery's revenue. The battery investment accounts for 81% of the total considered cost (which only includes battery- and cable-related costs, but does not consider the offshore wind investment cost). Another important observation is that the average battery degradation cost under the optimal solution for different locations is only 3.2% of the total cost. The revenue from the 1MWh battery adds up to $2.69 per MWh of wind energy which results in a maximum breakeven investment cost of $69.3 per kWh for the battery considering 10 years capital recovery period and 7% discount rate. This compares to current battery investment costs of $400-600 per kWh [11]. It will be useful to note here that our results indicate that the added value of 1MWh battery to 1MW wind asset through

energy- and capacity-market revenues is ~$2 and ~$0.6 per MWh of wind energy. Mills et al. [12] report added values of $1.9 and $0.6 MWh-wind for a 1.25MW/5MWh battery for 18MWs of offshore wind in NYISO. These numbers although not directly comparable, due in part to the difference in the wind energy output and price variations in different years, still gives some sense of how our numbers with advanced battery models compare to other existing studies.

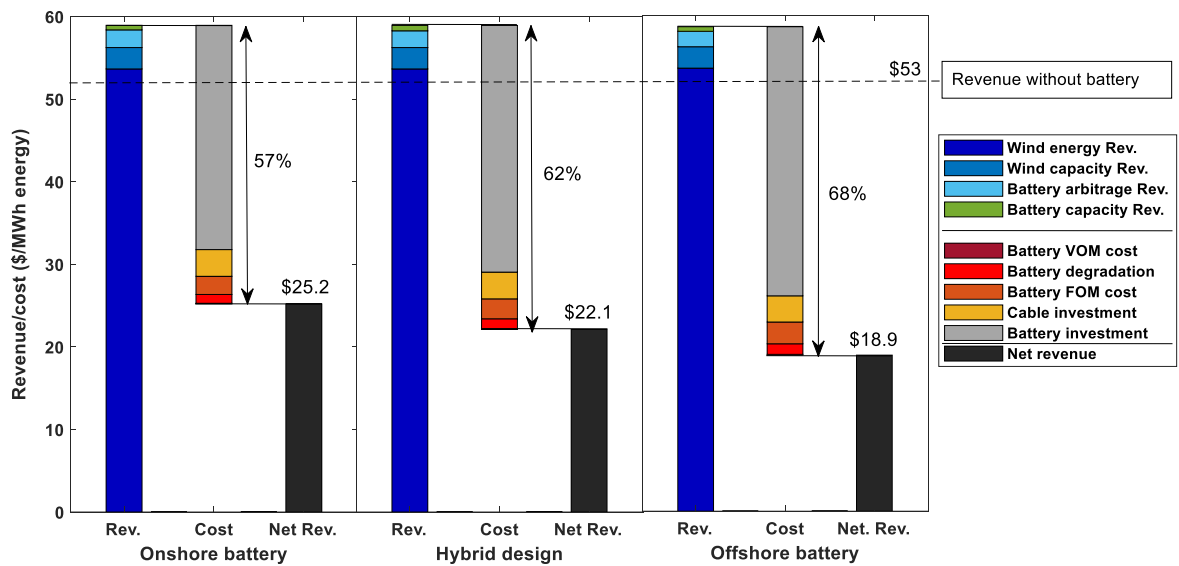

Fig. 9. Average (over 2010-2013 data) revenue and cost breakdown per MWh wind energy with different system configurations

Table 3. Average (over 2010-2013 data) revenue and cost breakdown and breakeven costs for different battery locations per MWh of wind energy

| $ | Without battery | Onshore battery | Hybrid design | Offshore battery |
|---|---|---|---|---|
| Wind energy revenue | 53.65 | 53.65 | 53.65 | 53.65 |
| Wind capacity value | 2.59 | 2.59 | 2.59 | 2.59 |
| Battery arbitrage revenue | 0 | 2.14 | 2.03 | 1.87 |
| Battery capacity value | 0 | 0.55 | 0.62 | 0.55 |
| **Total Revenue** | 56.24 | 58.93 | 58.89 | 58.66 |
| Battery Inv. cost | 0 | 27.15 | 29.87 | 32.58 |
| Cable Inv. cost | 3.24 | 3.24 | 3.24 | 3.17 |
| Battery fixed O&M cost | 0 | 2.20 | 2.41 | 2.64 |
| Capacity fade cost | 0 | 1.07 | 1.20 | 1.28 |
| Battery variable O&M cost | 0 | 0.08 | 0.09 | 0.10 |
| **Total cost** | 3.24 | 33.74 | 36.81 | 39.76 |
| **Net revenue** | 53.01 | 25.20 | 22.11 | 18.90 |
| Total battery revenue | - | 2.69 | 2.68 | 2.42 |
| **Breakeven cost of battery ($/kWh)** | - | **69.3** | **68.9** | **62.2** |

The BESS revenues and corresponding breakeven costs vary in different years due to energy and capacity market price variations as shown in Fig. 10. The BESS revenue is highest in 2013, while it is lowest for 2010 and 2011 for all battery locations. These results lead to breakeven cost interval of $50-95 across the four different years. Fig. 11 illustrates that capacity market revenues vary more than energy market revenues across the four years. The BESS energy and capacity market revenues breakdown indicate that although energy market revenue is higher in some years (e.g. 2012), the capacity market revenue can alter the total revenue result (e.g. 2013).

4-3. BESS SOC Operational Window

We also compare the BESS revenues for different useable SOC windows. Fig. 12 shows that widening the SOC window maximizes the battery capacity utilization and increases the gross revenue. On the other hand, deep cycling leads to excessive degradation and high associated costs, while the degradation rate is slower with narrow SOC windows. However, due to a high arbitrage opportunity in a specific hour, the increase in the revenue with higher DODs is larger than the increase in the degradation costs and therefore, the net revenue is higher for wider SOC windows.

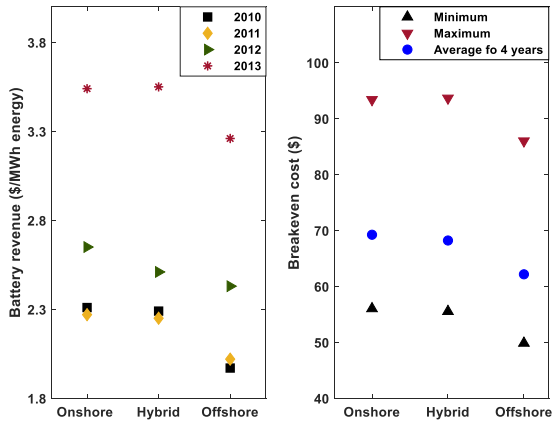

Fig. 10. BESS net revenue in different years and breakeven cost bound

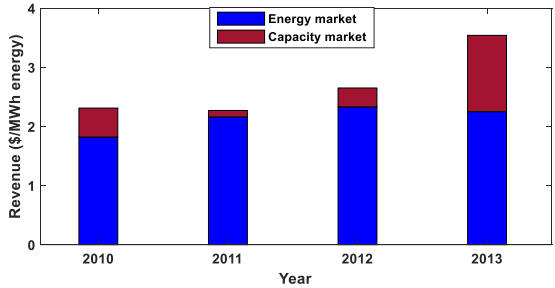

Fig. 11. Onshore BESS energy and capacity market revenue breakdown for different years

Operational results indicate that the energy throughput is higher for lower SOC windows in the optimal solution, although, it does not lead to higher revenues. Referring to the cycle life-DOD relationship in Fig. 3, in lower SOC windows the cycling degradation cost is low and the BESS is cycled even for small arbitrage opportunities to maximize the revenue, while with wider SOC windows the BESS is cycled only in higher arbitrage hours to avoid excessive degradation. For different SOC windows, 91-94% of the charged energy comes from the offshore wind and 6-9% is purchased from the grid. The annual roundtrip efficiency of the BESS for different SOC windows is 84-86%, with higher values for lower SOC windows.

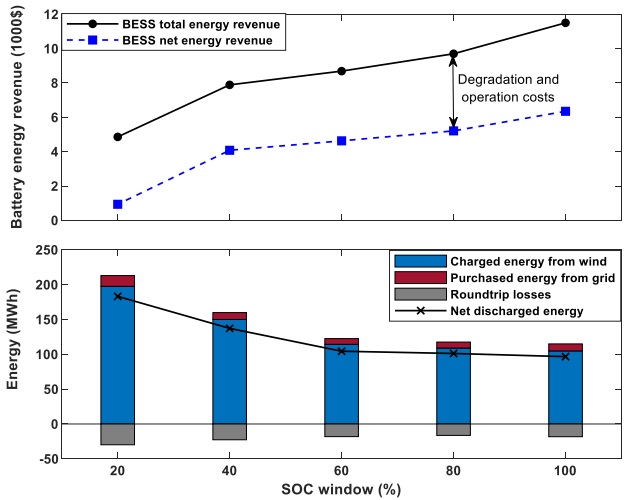

Fig. 12. Impact of BESS useable SOC window on its added revenue and cycling results for the 1MWh battery

### 4-4. Energy and Capacity Markets in NYISO

Finally, after exploring the BESS's technical configurations, we compare the impact of participation in different markets on the revenue of offshore wind connected BESS. Fig. 13 shows the revenue elements breakdown in NYISO for energy and capacity markets. These results are normalized for 1MW offshore wind and 1MWh onshore battery system. In the energy market, the average revenue across different years for wind is $198,000, compared to $7,800 for BESS. The variation in the wind and energy price data in different years has significant impact on the offshore wind revenue (±24%), while the BESS energy revenue has smaller variations (±8%).

In the capacity market, the BESS can be self-managed or managed by the ISO and these scenarios lead to different capacity revenues. The self-managed BESS has 40% lower capacity revenue. In both cases, the BESS capacity revenue is 64-78% lower than the wind capacity revenue. Also, due to the larger variations in the capacity prices in different years, both wind and BESS revenues have substantial variations in different years.

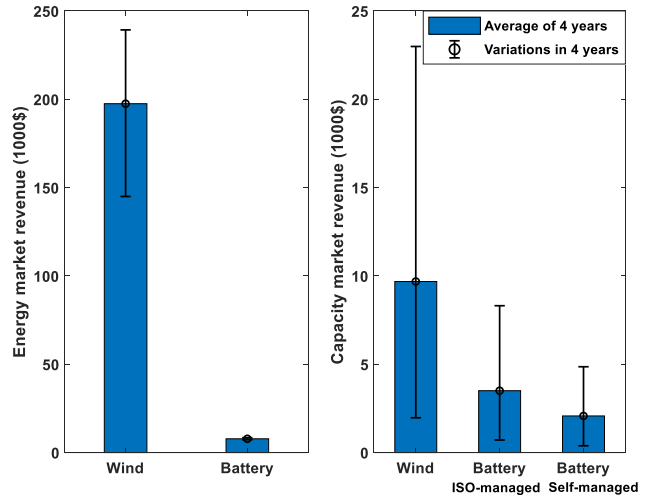

Fig. 13. Energy and capacity markets revenues per MW offshore wind for the 1 MWh battery (average and variation over four years)

### 5- SUMMARY AND CONCLUSIONS

We evaluated the economic viability of pairing BESSs together with offshore wind by quantifying revenues from participation in energy and capacity markets. Towards this end, we developed enhanced models of lithium-ion batteries that account for their dynamic efficiency as a function of the discharge power, power-limits as a function of the SOC, and cycle- and calendar-lives. The proposed representations can serve as a generic framework that captures the physical phenomena characterizing the behavior of BESSs in MILP format. Hence, this generic formulation can be directly implemented in a range of different power system optimization models. Using the proposed BESS formulation, we investigated multiple offshore wind and battery hybrid system designs to examine the impact of locating the battery offshore vs. onshore. We also explored the optimal BESS operation under different usable SOC windows to maximize energy- and capacity-market revenues for a selected location in NYISO using four years of wind and market price data.

We find that relying on simplistic BESS models adds substantial error to the battery revenue estimate. With our enhanced battery representations, net revenues can be increased by 29%. The analysis highlights the importance of using detailed and sophisticated BESS models by asset owners to accurately estimate the revenue potential and improve operational and planning decisions.

We also find that locating the BESS onshore and operating it within its full SOC window results in the greatest revenue potential. Cable losses, battery degradation, and the historical price profiles used in the study with specific periods of high arbitrage opportunities contribute to explain this outcome. We found that the energy revenue of the BESS is around 4% of the offshore wind revenue, while the BESS's capacity revenue is 22-36% of the offshore wind capacity revenue. Both energy- and capacity-market revenues, with the latter in particular, vary substantially in different years due to the weather and price uncertainties.

Overall, results underscore that BESS is still an expensive investment option. We find that the average breakeven price of storage in 4 years is $69 per kWh for the onshore battery (with $55-95 per kWh variation), which is well below current BESS costs. The presented numerical results are obviously limited to the specific inputs and assumptions of this study and should be interpreted accordingly. In future work, we plan to study a wider range of battery chemistries, including flow batteries. We also plan to consider other geographical locations, markets, and additional battery applications and revenue streams.

APPENDIX A

We present the basic cycling constraint of the BESS such as SOC and cycling power boundaries, SOC calculations, and power losses curves selection for different SOC levels in this section as follows:

$$P_d(t) \leq P_d^{Max} \times B(t) \quad \forall t \quad (A.1)$$

$$P_c(t) \leq P_c^{Max} \times (1 - B(t)) \quad \forall t \quad (A.2)$$

$$P_d^{loss}(t) \leq P_d^{Max} \times B(t) \quad \forall t \quad (A.3)$$

$$P_c^{loss}(t) \leq P_c^{Max} \times (1 - B(t)) \quad \forall t \quad (A.4)$$

$$C(t) = C_{int.} - P_d(t) + P_c(t) \quad t = 1 \quad (A.5)$$

$$S(t) = (C(t) + C_{int.})/2C_r \quad t = 1 \quad (A.6)$$

$$C(t) = C(t-1) - P_d(t) + P_c(t) \quad t \geq 2 \quad (A.7)$$

$$S(t) = (C(t) + C(t-1))/2C_r \quad t \geq 2 \quad (A.8)$$

Constraints (A.1) and (A.2) prevent charging and discharging to happen at the same time and (A.3) and (A.4) prevent from double counting of the losses in charging and discharging by a binary variable $B$. Constraints (A.5) and (A.7) calculate the BESS's energy level $C$ in MWh, and (A.6) and (A.8) compute the SOC level of the BESS in [0-1] scale at all time steps.

$$\alpha_d(t) \leq \sum_{l \in L} p_d(l) + M \times (1 - U(t,k)) \quad \forall t,k \quad (A.9)$$

$$\alpha_d(t) \geq \sum_{l \in L} p_d(l) - M \times (1 - U(t,k)) \quad \forall t,k \quad (A.10)$$

$$\alpha_c(t) \leq \sum_{n \in N} p_c(n) + M \times (1 - U(t,k)) \quad \forall t,k \quad (A.11)$$

$$\alpha_c(t) \geq \sum_{n \in N} p_c(n) - M \times (1 - U(t,k)) \quad \forall t,k \quad (A.12)$$

$$\sum_{k \in K} \sum_{l \in L} w_d(t,l,k) = P_d(t) \quad \forall t \quad (A.13)$$

$$\sum_{k \in K} \sum_{n \in N} w_c(t,n,k) = P_c(t) \quad \forall t \quad (A.14)$$

$$w_d(t,l,k) \leq p_d(l) \times U(t,k) \quad \forall t,l,k \quad (A.15)$$

$$w_c(t,n,k) \leq p_c(n) \times U(t,k) \quad \forall t,m,k \quad (A.16)$$

$$S(t) \leq \sum_{k \in K} \beta(k+1) \times U(t,k) \quad \forall t \quad (A.17)$$

$$S(t) \geq \sum_{k \in K} \beta(k) \times U(t,k) \quad \forall t \quad (A.18)$$

$$\sum_{k \in K} U(t,k) = 1 \quad \forall t \quad (A.19)$$

Constraints (A.9) and (A.10) calculate the maximum "discharging" power $\alpha_d$ for (16) with the power pieces $p_d$ from input data, using "Big M" method to select a SOC curve by a binary variable $U$. (A.11) and (A.12) repeats the same process for "charging" maximum power calculation.

Eqs. (A.13) and (A.14) sum up the power curve pieces to calculate the output power of the battery, while (A.15) and (A.16) activate the right power curve pieces for the losses calculation from the input loss-power curves for different SOCs. In (A.17) and (A.18), the binary variable for the SOC curve selection is activated for each time step and (A.19) makes sure that only one SOC curve is active at a time. The SOC bins are input to the model based on the experimental data of losses and maximum power limits and are defines as follows:

$$S_{dn} = \beta^1 \leq \beta^2 \leq \cdots \leq \beta^K = S_{up} \quad (A.20)$$

Therefore, to solve the problem for onshore BESS configuration as an example, the objective function (1) is subject to (2)-(5), (15)-(26) and (A.1)-(A.20).

ACKNOWLEDGMENTS

The authors would like to acknowledge Equinor ASA for supporting the project. In particular, the authors would like to thank Jan Henrik Borch and Børre Tore Børresen for their helpful comments and feedback. The authors thank Patrick R. Brown for his comments and help with data gathering. Acknowledgement is also due towards Francis O'Sullivan for his help in getting the project started. The authors declare no conflicting interests. The views expressed in the paper are solely of the authors.

## REFERENCES

[1] World Nuclear Association, "Comparison of Lifecycle Greenhouse Gas Emissions of Various Electricity Generation Sources," 2011.

[2] B. Y. Rack, "US seen front and center of global offshore wind expansion," *Mark. Intell.*, pp. 1–2, 2019.

[3] W. Musial, P. Beiter, P. Spitsen, J. Nunemaker, and V. Gevorgian, "2018 Offshore Wind Technologies Market Report, *US DOE Office of Energy Efficiency and Renewable Energy*, 2018

[4] P. Beiter *et al.*, "The Vineyard Wind Power Purchase Agreement : Insights for Estimating Costs of U . S . Offshore Wind Projects," no. February, 2019.

[5] Lazard, "Lazard's Levelized Cost of Energy Analysis - Version 12.0," November 2018

[6] L. Burdock, "The State Of The U . S . Offshore Wind Market," pp. 2–4, 2019.

[7] International Energy Agency, "Offshore Energy Outlook," 2018.

[8] S. Chu and A. Majumdar, "Opportunities and challenges for a sustainable energy future," *Nature*, vol. 488, no. 7411. pp. 294–303, 2012.

[9] "U.S. Energy Storage Monitor," *Wood Mackenzie Power Renewables*, no. March, 2019.

[10] B. Nykvist and M. Nilsson, "Rapidly falling costs of battery packs for electric vehicles," *Nat. cimate Chang.*, vol. 5, no. March, pp. 100–103, 2015.

[11] R. Fu, T. Remo, R. Margolis, R. Fu, T. Remo, and R. Margolis, "2018 U.S. Utility-Scale Photovoltaics- Plus -Energy Storage System Costs Benchmark," *Natl. Renew. Energy Lab.*, no. November, 2018.

[12] A. D. Mills, D. Millstein, S. Jeong, L. Lavin, R. Wiser, and M. Bolinger, "Estimating the Value of Offshore Wind Along the United States ' Eastern Coast Detailed Summary of Results," no. April, pp. 1–61, 2018.

[13] P. Beiter, W. Musial, L. Kilcher, M. Maness, and A. Smith, "An Assessment of the Economic Potential of Offshore Wind in the United States from 2015 to 2030," no. March, 2017.

[14] W. Kempton, S. McClellan, and D. Ozkan, "Massachusetts Offshore Wind Future Cost Study," *Univ. Delaware*, no. March, 2016.

[15] G. He *et al.*, "Optimal Bidding Strategy of Battery Storage in Power Markets Considering Performance-Based Regulation and Battery Cycle Life," *IEEE Trans. Smart Grid*, vol. 7, no. 5, pp. 2359–2367, 2016.

[16] K. Bradbury, L. Pratson, and D. Patiño-Echeverri, "Economic viability of energy storage systems based on price arbitrage potential in real-time U.S. electricity markets," *Appl. Energy*, 2014.

[17] D. M. Davies *et al.*, "Combined economic and technological evaluation of battery energy storage for grid applications," *Nat. Energy*, 2018.

[18] F. Wankmüller, P. R. Thimmapuram, K. G. Gallagher, and A. Botterud, "Impact of battery degradation on energy arbitrage revenue of grid-level energy storage," *J. Energy Storage*, vol. 10, pp. 56–66, 2017.

[19] A. Sakti *et al.*, "Enhanced representations of lithium-ion batteries in power systems models and their effect on the valuation of energy arbitrage applications," *J. Power Sources*, vol. 342, pp. 279–291, 2017.

[20] M. Jafari, K. Rodby, J. L. Barton, F. Brushett, and A. Botterud, "Improved Energy Arbitrage Optimization with Detailed Flow Battery Characterization," in *IEEE PES General Meeting*, 2019.

[21] M. Jafari, K. Khan, and L. Gauchia, "Deterministic models of Li-ion battery aging: It is a matter of scale," *J. Energy Storage*, vol. 20, no. May, pp. 67–77, 2018.

[22] M. Jafari, A. Gauchia, S. Zhao, K. Zhang, and L. Gauchia, "Electric Vehicle Battery Cycle Aging Evaluation in Real-World Daily Driving and Vehicle-to-Grid Services," *IEEE Trans. Transp. Electrif.*, vol. 7782, no. c, pp. 1–1, 2017.

[23] B. Xu, A. Oudalov, A. Ulbig, G. Andersson, and D. S. Kirschen, "Modeling of Lithium-Ion Battery Degradation for Cell Life Assessment," *IEEE Trans. Smart Grid*, vol. 9, no. 2, pp. 1131–1140, 2018.

[24] Z. T. Smith, "Capacity Market Rules for Energy Storage Resources," *New York Indep. Syst. Oper.*, 2018.

[25] Panasonic Battery, "Sanyo E 18650 Li-ion Battery (UR18650E) data sheet," 2012.

[26] W. Cole and A. W. Frazier, "Cost Projections for Utility-Scale Battery Storage," *Natl. Renew. Energy Lab.*, no. June, 2019.

[27] P. Beiter, W. Musial, A. Smith, L. Kilcher, R. Damiani, M. Maness, S. Sirnivas, T. Stehly, V. Gevorgian, M. Mooney, and G. Scott, "A Spatial-Economic Cost Reduction Pathway Analysis for U.S. Offshore Wind Energy Development from 2015-2030" September 2016

[28] F. Windmonitor, "Heatmap of installed offshore wind turbine locations by water depth and distance from the coast," http://windmonitor.iee.fraunhofer.de/windmonitor_en/4_Offshore/2_technik/2_Kuestenentfernung_und_Wassertiefe/, accessed 22 Jan 2020

[29] C. Byers, T. Levin, and A. Botterud, "Capacity market design and renewable energy : Performance incentives , qualifying capacity , and demand curves," *Electr. J.*, vol. 31, no. 1, pp. 65–74, 2018.